\documentclass[pdflatex,sn-mathphys-num]{sn-jnl}

\usepackage{graphicx}%
\usepackage{float}%

\usepackage{multirow}%
\usepackage{amsmath,amssymb,amsfonts}%
\usepackage{amsthm}%
\usepackage{mathrsfs}%
\usepackage[title]{appendix}%
\usepackage{xcolor}%
\usepackage{textcomp}%
\usepackage{manyfoot}%
\usepackage{booktabs}%
\usepackage{algorithm}%
\usepackage{algorithmicx}%
\usepackage{algpseudocode}%
\usepackage{listings}%

\theoremstyle{thmstyleone}%
\theoremstyle{thmstyletwo}%

\theoremstyle{thmstylethree}%

\begin{document}

\title[CancerZigZag]{CancerZigZag: Iterative Seed-Anchored Diffusion for Generative Modeling of Single-Cell State Transitions}


\author*[1]{\fnm{Johannes} \sur{Schlüter}}\email{j.schlueter@uni-bielefeld.de}


\author*[1]{\fnm{Alexander} \sur{Schönhuth}}\email{aschoen@cebitec.uni-bielefeld.de}

\affil*[1]{\orgdiv{Faculty of Technology}, \orgname{Bielefeld University}, \orgaddress{\street{Universitätsstraße 25}, \city{Bielefeld}, \postcode{33615}, \state{NRW}, \country{Germany}}}



\abstract{Single-cell cancer datasets are predominantly cross-sectional and rarely provide paired or longitudinal observations linking individual healthy-like cells to tumor-associated states. This limits the study of state-associated variation at the level of individual starting cells. We introduce CancerZigZag, a seed-initialized diffusion-based framework for exploratory generation of tumor-associated single-cell candidate clouds from unpaired epithelial cell populations. For each cancer context, CancerZigZag trains a diffusion model exclusively on tumor-derived epithelial cells and applies repeated partial latent-space perturbation and reverse diffusion to held-out healthy-like seed cells. The method generates stochastic candidate clouds without paired healthy--tumor measurements or classifier guidance during generation.

We applied CancerZigZag to colorectal, breast, lung, and renal cell carcinoma contexts and explored parameter landscapes defined by perturbation depth and the number of ZigZag cycles. Across the reported operating configurations, candidate clouds contained outputs classified toward held-out tumor-derived reference populations for each evaluated seed. Residual seed-dependent organization varied substantially across contexts, with the clearest struc ture observed in colorectal cancer, more modest organization in lung cancer, and limited cloud-level structure in the reported breast and renal cell carcinoma settings. Representative candidates additionally showed directional concordance with transcriptional shifts observed between held-out healthy-like and tumor-derived reference populations.

CancerZigZag is therefore not interpreted as a model of deterministic healthy-to-tumor transformation or cellular progression. Instead, it provides a reference-informed framework for exploring tumor-associated candidate distributions from unpaired healthy-like seeds and for quantifying the context-dependent relationship between tumor-associated displacement and residual seed dependence.}




\maketitle
\section{Introduction}\label{sec1}

\begin{figure}[H]
\centering
\includegraphics[width=\linewidth]{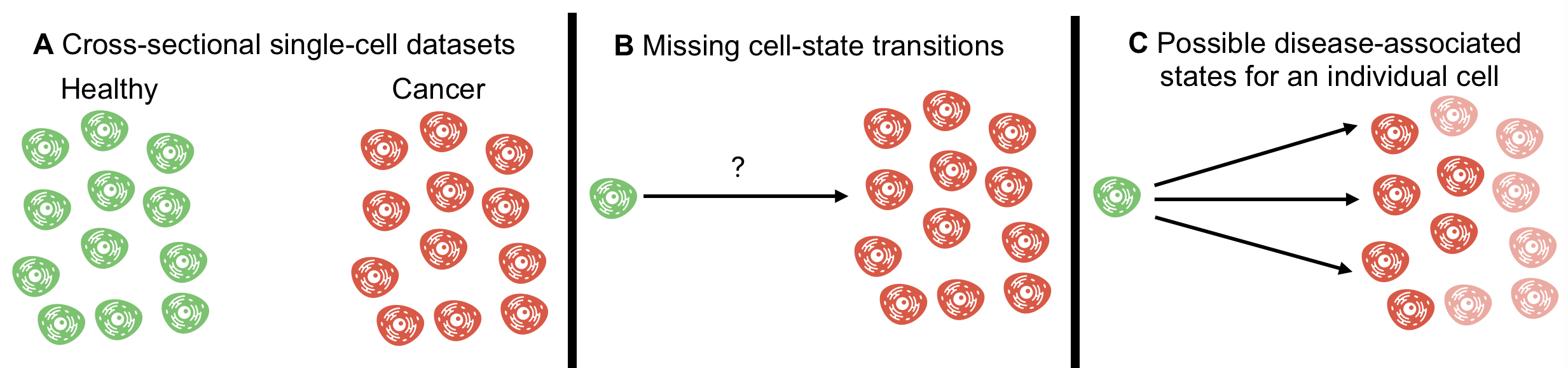}
\caption{Overview of the motivating problem. \textbf{A}: Single-cell cancer datasets are typically cross-sectional and contain heterogeneous healthy-like, non-malignant, and malignant cell populations, but rarely provide paired healthy--cancer measurements for individual cells. \textbf{B}: Cell-state relationships between healthy-like cells and potential cancer-associated states are therefore unobserved at the level of individual cells. \textbf{C}: In the absence of paired or longitudinal measurements, the goal is not to predict a unique future cancer cell for each healthy-like seed, but to explore candidate states exhibiting tumor-associated expression characteristics under explicit evaluation criteria.}
\label{fig:problem}
\end{figure}

Single-cell transcriptomic studies have enabled detailed characterization of healthy and diseased tissues across diverse biological contexts~\cite{Regev2017,Stuart2019}. However, most available single-cell cancer datasets are cross-sectional and comprise cells collected from different individuals rather than paired or longitudinal observations of state change, as exemplified by large pan-cancer tumor--normal atlas resources~\cite{Kang2024}. In the absence of paired measurements, lineage information, or longitudinal sampling, the relationship between an individual healthy-like epithelial cell and potential tumor-associated states is not directly observed, and a unique future tumor state cannot be inferred from snapshot data alone~\cite{Weinreb2018,Tritschler2019}. Computational models may instead be used to explore candidate states exhibiting tumor-associated expression characteristics under explicit evaluation criteria (Figure~\ref{fig:problem}).

Existing computational approaches address related aspects of cell-state modelling but do not directly target the setting considered here. Trajectory inference and pseudotime methods organize observed cells along inferred continua or directional structures, but do not generate new candidate expression states for individual starting cells~\cite{Trapnell2014,Haghverdi2016,Saelens2019}. Perturbation-response and conditional generative models can predict condition-associated cellular responses, but generally rely on explicit condition labels, perturbation annotations, or supervised contrasts~\cite{Lotfollahi2019,Lotfollahi2023CPA}. Diffusion-based approaches for single-cell data have demonstrated the ability to model expression distributions, generate synthetic profiles, and predict condition- or development-associated cellular states~\cite{Luo2024scDiffusion,Dong2026scRDiT,He2026Squidiff}; however, iterative candidate-cloud exploration initialized from individual healthy-like epithelial cells in an unpaired tumor setting remains insufficiently studied. Thus, a distinct methodological question is whether a generative model learned from tumor-derived cells can produce tumor-associated candidate clouds from unpaired healthy-like seed cells while retaining measurable dependence on the originating seed in at least some biological contexts.

We therefore developed CancerZigZag, a seed-initialized generative framework based on repeated partial noising and reverse diffusion in a tumor-trained latent expression landscape. Unlike conditional or classifier-guided diffusion strategies~\cite{Dhariwal2021,Ho2022CFG,Luo2024scDiffusion}, CancerZigZag does not optimize generated samples toward an external tumor objective during generation. Instead, healthy-like epithelial cells provide initial latent seed states, while a diffusion model trained only on tumor-derived epithelial cells defines the generative operator used to explore nearby and progressively displaced candidate states. Repeating this operation yields a stochastic candidate cloud for each seed rather than a single predicted tumor counterpart.

CancerZigZag is evaluated as a candidate-cloud framework rather than as a deterministic state-translation model. The complete stochastic cloud generated from each healthy-like seed constitutes the primary model output and is used to assess tumor-associated state acquisition and residual seed-dependent organization. Representative candidates selected under a predefined tumor-association and seed-distance rule provide secondary summaries, while generated paths are analysed descriptively to characterize expression programmes represented among successful outputs. These analyses are not interpreted as temporal progression, lineage reconstruction, or deterministic transformation of healthy-like cells into tumor cells.

\begin{figure}
\centering
\includegraphics[width=\linewidth]{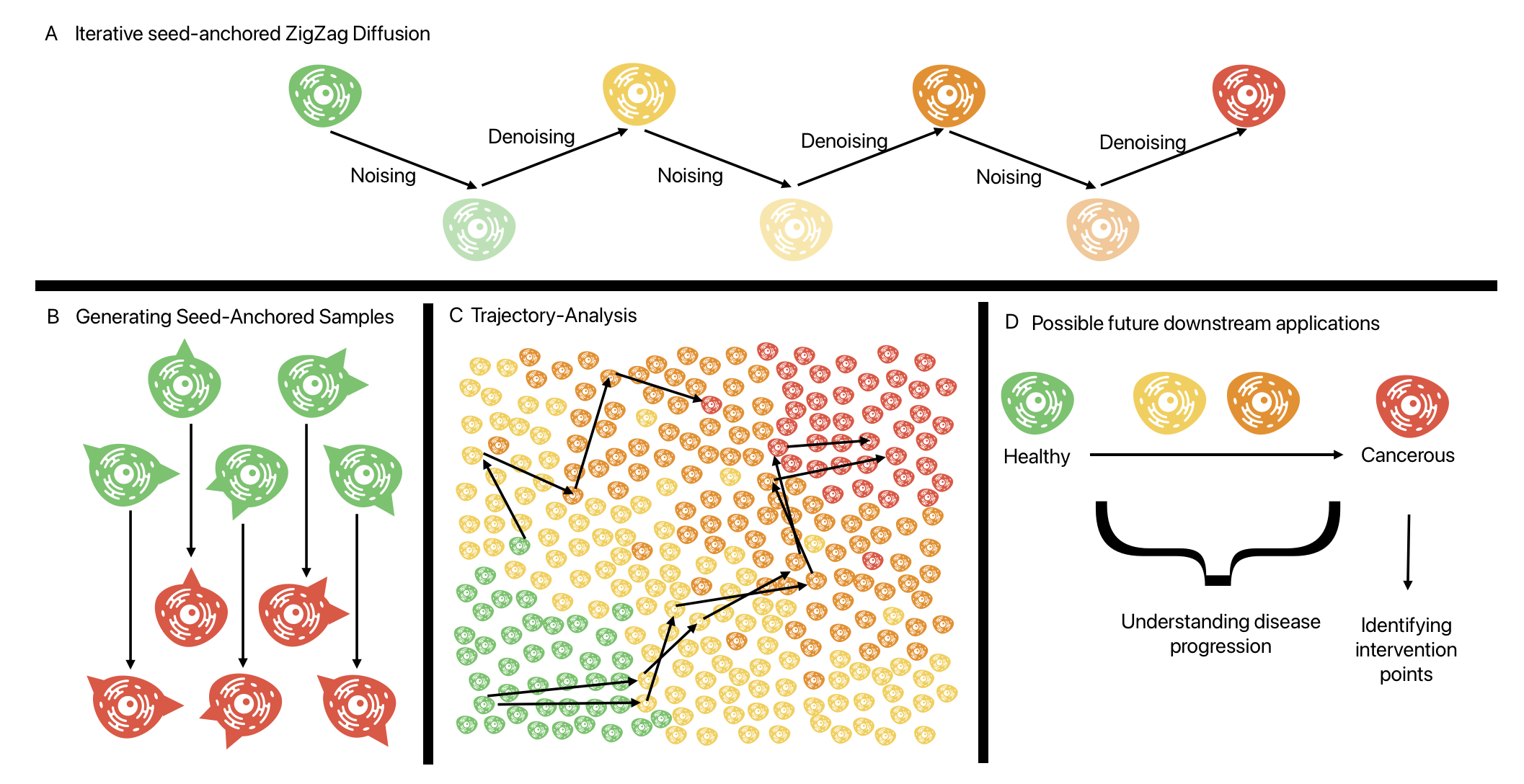}
\caption{\textbf{A} CancerZigZag applies repeated partial noising and tumor-trained reverse diffusion to a latent representation initialized from a healthy-like epithelial seed cell. \textbf{B} Repeated stochastic realizations produce a candidate cloud for each seed rather than a single predicted tumor counterpart. \textbf{C} Generated paths can be inspected descriptively to characterize movement through the learned tumor-associated latent landscape; they are not interpreted as temporal or causal trajectories. \textbf{D} Candidate outputs can be evaluated for tumor-associated scoring, residual seed-dependent structure, and directional concordance with transcriptional programmes observed in held-out tumor-derived reference cells.}
\label{fig:main_figure}
\end{figure}

We applied CancerZigZag to four epithelial cancer contexts and explored how the number of ZigZag cycles and the perturbation depth affect the relationship between tumor-associated candidate generation and residual seed dependence. Because configurations were identified from reference-informed parameter landscapes, the reported analyses are interpreted as exploratory operating-regime characterization rather than as independently preselected predictive-performance estimates. Seed dependence was assessed using complementary summaries, including representative-candidate anchoring and candidate-cloud organization after removal of the dominant healthy-to-tumor reference direction.

Across the reported operating configurations, CancerZigZag candidate clouds contained classifier-defined tumor-associated candidates for each evaluated held-out seed. Residual seed-dependent structure was context-dependent: it was most evident in CRC, more modest in LC, and limited at the cloud level in the reported BRCA and RCC settings. Representative candidates additionally showed directional agreement with tumor-associated transcriptional shifts observed in held-out reference populations. These findings support CancerZigZag as an exploratory framework for generating and characterizing tumor-associated candidate clouds from unpaired healthy-like seed cells, while highlighting that tumor-associated state acquisition and residual seed dependence represent distinct, context-dependent properties of the generated outputs.

\section{Methods}\label{sec11}

\subsection{Overview of the CancerZigZag framework}

\begin{figure}[H]
    \centering
    \includegraphics[width=\linewidth]{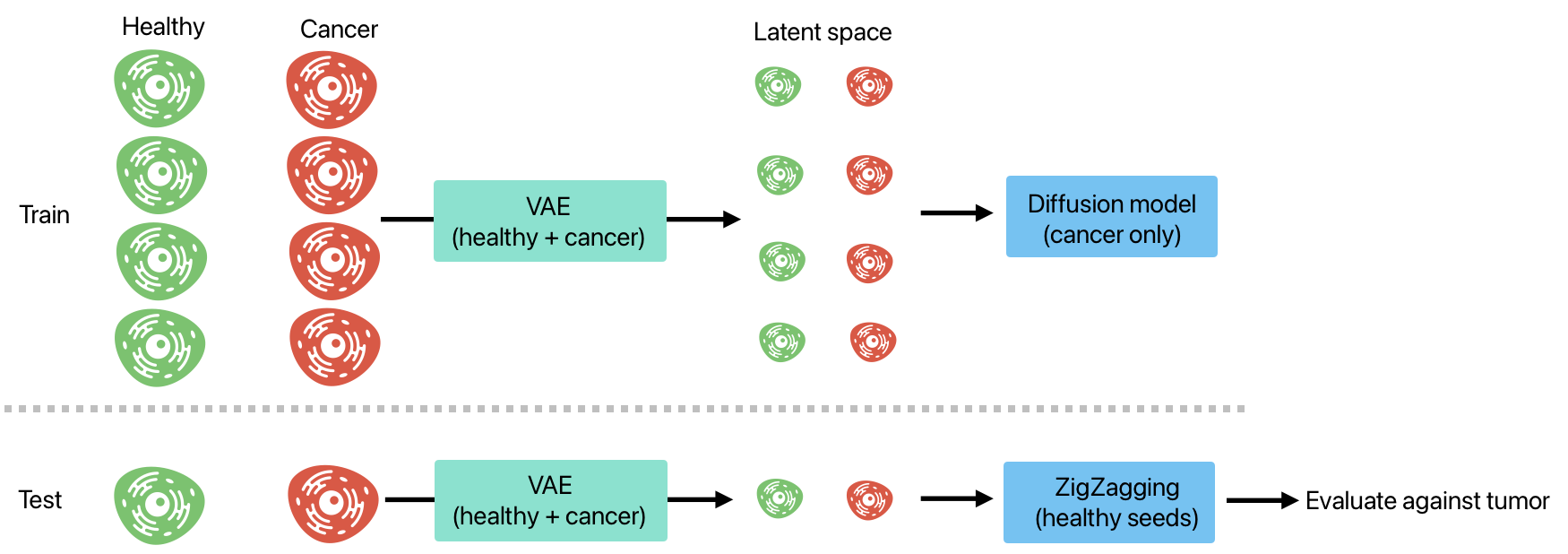}
    \caption{We apply a strict separation of training and test data (See Table~\ref{tab:data_usage_2x2}): First, the VAE is finetuned on training samples. Then, we train the diffusion model on tumor cells from the training cohort only. For testing, healthy-like and tumor cells from the held-out cohort are encoded into the fixed VAE latent space. CancerZigZag starts from healthy-like latent seed cells and generates tumor-like candidate states using the tumor-trained diffusion model. Generated candidates are evaluated against held-out test tumor references. Training, generation, and evaluation are performed independently for each dataset to assess generalizability across cancer contexts.}
    \label{fig:workflow}
\end{figure}

CancerZigZag generates stochastic candidate clouds from healthy-like epithelial seed cells by repeatedly applying partial latent-space perturbation and reverse diffusion using a model trained on tumor-derived epithelial cells. Complete candidate clouds constitute the primary output; representative candidates and generated paths are used only for secondary descriptive summaries. The complete workflow and the separation of training and held-out evaluation cohorts are summarized in Fig.~\ref{fig:workflow} and Table~\ref{tab:data_usage_2x2}.

\subsection{Single-cell transcriptomic dataset}

We used single-cell RNA-sequencing data from the pan-cancer tumor--normal atlas by Kang et al.~\cite{Kang2024}, obtained from the associated Zenodo repository (\url{https://doi.org/10.5281/zenodo.10651059}). The original atlas integrates post-quality-control single-cell transcriptomes from tumor and normal tissue samples across 30 cancer types. For the present study, we used the epithelial-cell expression object provided in the atlas resource and analysed four cancer contexts separately: breast cancer (BRCA), colorectal cancer (CRC), lung cancer (LC), and renal cell carcinoma (RCC). Spatial transcriptomics data and NMF-derived cell-state representations provided with the atlas resource were not used for CancerZigZag model training or candidate generation.

\subsection{Single-cell data preprocessing}

For each cancer context, epithelial cells were selected according to the cancer-type annotations provided in the source atlas. Cells were then restricted to epithelial populations classified by the atlas as either non-malignant or malignant on the basis of inferred copy-number variation profiles~\cite{Kang2024}. Cells classified as non-malignant were used as healthy-like cells, whereas cells classified as malignant were used as tumor-derived cells.

Expression profiles were aligned to the gene feature space required by the pretrained scDiffusion representation model~\cite{Luo2024scDiffusion}. Genes not observed in a given cancer-context-specific expression matrix were assigned zero expression during alignment, yielding a common feature space across all analysed contexts. Where raw count data were available in the source atlas, these were used as input prior to normalization and log transformation; otherwise, the expression values supplied in the processed atlas object were retained.

Healthy-like and tumor-derived epithelial cells were treated as unpaired source and reference populations. No cell-level correspondences, matched healthy--tumor pairs, longitudinal observations, lineage relationships, or patient-specific transition labels were used during preprocessing, model training, candidate generation, or evaluation.

\subsection{Patient-wise data splitting and evaluation protocol}

To assess generalization across patients and prevent patient-level information leakage, train--test splitting was performed separately for each cancer context at the patient level. Approximately 20\% of patients were assigned to a held-out evaluation cohort. Where possible, the held-out cohort included both healthy-like and tumor-derived epithelial cells to enable seed-initialized generation and reference-based post hoc evaluation within the same cancer context. Training and evaluation cohorts were disjoint, and no patient contributed cells to both partitions.

The VAE was fine-tuned using cells from the training cohort only. The diffusion model was subsequently trained exclusively on tumor-derived epithelial cells from the same training cohort. Healthy-like epithelial cells from the held-out evaluation cohort were used as seed cells for CancerZigZag generation, whereas healthy-like and tumor-derived epithelial cells from the held-out evaluation cohort served as biological reference populations for post hoc evaluation of generated candidates. No cells from held-out patients were used for VAE fine-tuning or diffusion-model training (See Table~\ref{tab:data_usage_2x2}).

\subsection{Variational autoencoder}

To obtain a low-dimensional latent representation of the aligned single-cell expression profiles, we used the pretrained variational autoencoder (VAE) distributed with the scDiffusion implementation, which builds on the scimilarity representation framework~\cite{Luo2024scDiffusion,Heimberg2025SCimilarity}. For each cancer context, the pretrained VAE model was fine-tuned separately on the corresponding patient-wise training partition, comprising the retained epithelial cells from that cancer context.

Prior to VAE fine-tuning and subsequent latent-space encoding, aligned expression matrices were normalized to a total count of $10^{4}$ per cell and log-transformed using $\log(1+x)$. The VAE was fine-tuned using the latent dimensionality of 128 adopted in the original scDiffusion framework~\cite{Luo2024scDiffusion}, with the Adam optimizer, a learning rate of $10^{-4}$, 40 epochs, and a maximum of 100,000 training cells per cancer context.
Fine-tuning minimized a reconstruction objective with latent-space regularization:
\[
\mathcal{L}_{\mathrm{VAE}} =
\mathrm{MSE}(x,\hat{x}) +
\beta \cdot \frac{1}{2}\mathbb{E}\left[z^{2}\right],
\]
where $x$ denotes the normalized log-transformed expression vector, $z$ its latent representation, $\hat{x}$ its reconstruction, and $\beta = 0.05$.

Following fine-tuning, the cancer-context-specific VAE was kept fixed for all subsequent diffusion model training and CancerZigZag generation steps. Tumor-derived epithelial cells from the respective training partition were encoded into the fine-tuned latent space for diffusion model training. During CancerZigZag generation, held-out healthy-like epithelial seed cells were encoded using the same fixed encoder, and generated latent states were decoded using the corresponding fixed decoder for downstream expression-space analyses.

\subsection{Diffusion model training}

The diffusion component of CancerZigZag was based on denoising diffusion probabilistic modelling~\cite{Ho2020DDPM}, as implemented for single-cell expression modelling in scDiffusion~\cite{Luo2024scDiffusion}. We used this framework to learn cancer-context-specific tumor-associated latent distributions and subsequently applied the trained reverse diffusion process within the iterative CancerZigZag generation procedure.

For each cancer context, a diffusion model was trained exclusively on latent representations of tumor-derived epithelial cells from the corresponding patient-wise training partition. These latent representations were obtained using the fixed, cancer-context-specific fine-tuned VAE encoder. During diffusion model training, additive Gaussian noise was applied to tumor-cell latent representations at randomly sampled diffusion time steps, and the model was optimized to predict the added noise using a mean-squared-error objective.

No healthy-like epithelial cells and no cells from held-out test patients were used for diffusion model training. Consequently, the diffusion model learns a tumor-associated latent distribution without access to paired healthy--tumor correspondences, longitudinal measurements, or lineage information. Healthy-like epithelial cells enter the workflow only at generation time, where their latent representations serve as initial seed states for repeated partial noising and reverse diffusion using the tumor-trained model.

\subsection{CancerZigZag generation procedure}

For each healthy-like epithelial seed cell $x_i$, an initial latent representation
was obtained using the fixed, cancer-context-specific fine-tuned VAE encoder,
\[
z_{i,b}^{(0)} = E_{\phi}(x_i),
\qquad b = 1,\ldots,B,
\]
where $B$ denotes the number of independent stochastic realizations generated
from each seed. All realizations are initialized from the same encoded seed
state, but evolve independently through newly sampled diffusion noise.

CancerZigZag generation consists of $r$ consecutive ZigZag cycles. In each
cycle, the current latent state is partially perturbed to a predefined diffusion
time index $t$ and is subsequently denoised through a reverse diffusion pass
using the diffusion model trained exclusively on tumor-derived epithelial cells
from the corresponding training cohort. Accordingly, $r$ denotes the number of
complete partial-noising/reverse-diffusion cycles, whereas $t$ controls the
perturbation depth applied within each cycle.

Let $z_{i,b}^{(k-1)}$ denote the latent state of realization $b$ from seed $i$
before ZigZag cycle $k$, where $k=1,\ldots,r$. Following the forward diffusion formulation of denoising diffusion probabilistic models, as adopted in scDiffusion~\cite{Ho2020DDPM,Luo2024scDiffusion}, a fully perturbed proposal at diffusion time $t$ is obtained as
\[
\tilde{z}_{i,b,t}^{(k)}
=
\sqrt{\bar{\alpha}_{t}}\,z_{i,b}^{(k-1)}
+
\sqrt{1-\bar{\alpha}_{t}}\,\epsilon_{i,b}^{(k)},
\qquad
\epsilon_{i,b}^{(k)} \sim \mathcal{N}(0,I),
\]

where $\bar{\alpha}_{t}$ denotes the cumulative diffusion noise schedule.

CancerZigZag applies a damping factor $\eta$ to the forward perturbation.
Rather than replacing the current latent state by the fully perturbed proposal,
the state entering reverse diffusion is defined as
\[
u_{i,b,t}^{(k)}
=
z_{i,b}^{(k-1)}
+
\eta
\left(
\tilde{z}_{i,b,t}^{(k)}
-
z_{i,b}^{(k-1)}
\right),
\qquad
0 < \eta \leq 1.
\]
Thus, $\eta=1$ corresponds to a full forward-diffusion perturbation, whereas
values $\eta<1$ retain a larger contribution of the current latent state during
each perturbation step.

Starting from $u_{i,b,t}^{(k)}$, the tumor-trained diffusion model performs a
reverse diffusion pass. At each reverse step, let
$\tilde{v}_{i,b,j-1}^{(k)}$ denote the sample proposed by the learned reverse
transition from the current reverse-process state $v_{i,b,j}^{(k)}$. The same
damping factor $\eta$ is applied to each reverse update:
\[
v_{i,b,j-1}^{(k)}
=
v_{i,b,j}^{(k)}
+
\eta
\left(
\tilde{v}_{i,b,j-1}^{(k)}
-
v_{i,b,j}^{(k)}
\right).
\]
After completion of the reverse diffusion pass, the output of cycle $k$ is
defined as
\[
z_{i,b}^{(k)} = v_{i,b,0}^{(k)}.
\]
This state becomes the starting point for the next ZigZag cycle. After $r$
cycles, the final latent candidate is
\[
\hat{z}_{i,b} = z_{i,b}^{(r)},
\]
and the corresponding expression-space candidate is obtained using the fixed
VAE decoder,
\[
\hat{x}_{i,b} = D_{\phi}(\hat{z}_{i,b}).
\]

For each seed $x_i$, CancerZigZag therefore generates a stochastic candidate
cloud
\[
\mathcal{C}_i =
\left\{
\hat{x}_{i,1},
\hat{x}_{i,2},
\ldots,
\hat{x}_{i,B}
\right\}.
\]
The candidate cloud, rather than a deterministic one-to-one healthy-to-tumor
mapping, constitutes the primary generative output of CancerZigZag.

For the analyses reported in this study, CancerZigZag generation was performed without classifier-based or centroid-based guidance of the type used in guided diffusion approaches~\cite{Dhariwal2021,Ho2022CFG,Luo2024scDiffusion}. Tumor-associated scores were computed only after candidate generation and did not influence the generative process.

\begin{table}[t]
\centering
\caption{\textbf{Overview of data usage across training, generation, and evaluation.}
The diffusion model is trained exclusively on tumor-derived epithelial cells
from the training cohort. Healthy-like epithelial cells from the held-out test
cohort are used as seed cells during generation, and tumor-derived epithelial
cells from the held-out evaluation cohort are used as evaluation references.}
\label{tab:data_usage_2x2}
\begin{tabular}{lcc}
\toprule
 & \textbf{Training cohort} & \textbf{held-out evaluation cohort} \\
\midrule
\textbf{Tumor-derived cells}
& Diffusion model training
& Evaluation references \\
\textbf{Healthy-like cells}
& Not used for diffusion training
& Seed cells for CancerZigZag generation \\
\bottomrule
\end{tabular}
\end{table}

\subsection{CancerZigZag parameter grid and evaluation configurations}

We evaluated the effect of exploration depth and repeated ZigZag cycling using
a systematic parameter grid separately for each of the four cancer contexts:
BRCA, CRC, LC, and RCC. The number of consecutive ZigZag cycles was varied as
\[
r \in \{1,10,25,50,75,100\},
\]
and the perturbation diffusion time index was varied as
\[
t \in \{1,10,25,50,75,100\}.
\]
This resulted in $36$ CancerZigZag configurations per cancer context and
$144$ evaluated cancer-context/configuration combinations in total.

The parameter $r$ controls how often partial perturbation and tumor-trained
reverse diffusion are applied sequentially, whereas $t$ controls the
perturbation depth within each individual cycle. The damping factor $\eta$ was
held fixed at $\eta = 0.1$ across the reported grid analyses.

For each evaluated configuration, $B=100$ independent stochastic candidate
realizations were generated per healthy-like seed. Holding the candidate
sampling budget fixed across the parameter grid allowed differences between
configurations to be attributed to the effects of $r$ and $t$ rather than to
differences in the number of generated candidates.

The configuration \(r = 1\) represents a single partial-noising/reverse-diffusion cycle and was included as a direct ablation of repeated ZigZag cycling. Configurations with \(r > 1\) evaluate the effect of iterative exploration through the tumor-trained latent generative landscape. Across the systematic parameter grid, generated outputs were evaluated using classifier-based tumor-association scores and seed-structure summaries. Additional biological programme analyses and targeted falsification analyses were performed for the reported main configurations where explicitly stated.

Tumor-derived epithelial cells from the held-out evaluation cohort were not used for VAE fine-tuning, diffusion-model training, or CancerZigZag generation. They were used only as biological reference cells for post hoc evaluation of generated candidates and for exploratory identification of reported operating configurations from the \(r \times t\) parameter landscapes. Accordingly, reported configuration-specific summaries are interpreted as reference-informed candidate-discovery results rather than as estimates of independently preselected predictive performance.

\subsection{Candidate generation and success definition}

For each healthy-like seed cell, CancerZigZag generates a stochastic candidate cloud consisting of \(B\) independently sampled candidate states. The complete candidate cloud constitutes the primary output of the method and was evaluated with respect to tumor-associated state acquisition, residual seed-conditioned structure and complementary transcriptional-shift concordance of representative candidates.

For analyses requiring a binary designation of tumor-associated candidates,
candidates with a classifier-estimated tumor probability of at least $0.7$
were designated as successful candidates under the predefined operational
criterion described below. This success definition was used to quantify the
presence and diversity of tumor-associated candidate states within each cloud
and did not influence CancerZigZag generation.

Representative single-candidate summaries were computed only where explicitly
reported, for example to compare predefined selection rules or to summarize
one candidate per seed. For such analyses, candidates satisfying the
classifier-estimated tumor-probability threshold were ranked by their
expression-space distance to the originating seed; if no candidate reached the
threshold, the candidate with maximal tumor logit was retained as a fallback.
These representative-candidate analyses are complementary to, rather than a
replacement for, the primary candidate-cloud evaluation.

For expression-space evaluation, we introduced a gene-wise detection-rate matching procedure that postprocessed decoded profiles relative to the held-out tumour-derived epithelial reference population. This projection was used to reduce mismatches in gene-wise sparsity and detection characteristics between VAE-decoded profiles and observed single-cell expression profiles, which are relevant technical features of scRNA-seq data~\cite{Hafemeister2019,Svensson2020}. Accordingly, expression-space classifier scores, seed-distance summaries, residual expression-space structure, and transcriptional-shift analyses are interpreted as reference-conditioned evaluations of sparsity-adjusted decoded profiles. The projection was applied only after latent candidate generation and did not influence VAE fine-tuning, diffusion-model training, latent-space CancerZigZag generation, or stochastic candidate exploration.

\subsection{Tumor-associated state evaluation}

Tumor-associated properties of generated candidate states were evaluated post hoc using a reference classifier. VAE-decoded healthy-like and tumor-derived epithelial reference cells from the corresponding held-out evaluation cohort were used to fit a binary evaluation classifier. These reference profiles entered classifier fitting as decoded representations without the candidate-specific gene-wise sparsity projection described above. Expression profiles were standardized and reduced to 50 principal components before fitting a class-balanced logistic regression classifier.

The fitted classifier was subsequently applied to generated candidate profiles after gene-wise sparsity projection guided by detection rates in the held-out tumor-derived epithelial reference population. Because this score quantifies agreement of postprocessed generated candidates with observed held-out reference populations, it is interpreted as a post hoc reference-conditioned tumor-association score rather than as an independently predicted biological probability of malignancy.

Let \(f(\hat{x}_{i,b})\) denote the decision function of the fitted reference classifier for generated candidate \(\hat{x}_{i,b}\). We define the tumor logit as
\[
s_{\mathrm{logit}}(\hat{x}_{i,b}) = f(\hat{x}_{i,b}),
\]
and the corresponding classifier-estimated tumor probability as
\[
s_{\mathrm{prob}}(\hat{x}_{i,b})
=
P_{\mathrm{clf}}\!\left(
y=\mathrm{tumor}
\mid
\hat{x}_{i,b}
\right).
\]
Higher values indicate stronger post hoc classification toward the held-out tumor-derived reference population.

For analyses requiring a binary designation of successful tumor-associated candidates, we used the predefined operational threshold
\[
s_{\mathrm{prob}}(\hat{x}_{i,b}) \geq 0.7.
\]
This threshold was used only for downstream evaluation and representative-candidate selection and did not influence CancerZigZag generation.

\subsection{Seed proximity and residual seed-conditioned structure}

CancerZigZag is designed to generate stochastic candidate clouds from healthy-like seed cells rather than to preserve exact cellular identity. We therefore evaluated seed dependence using projected expression-space seed distances and residual seed-structure metrics rather than interpreting candidate states as identity-preserving transformations.

For expression-space summaries, let \(\tilde{x}_{i,b}^{(T)}\) denote the decoded generated candidate \(\hat{x}_{i,b}\) after gene-wise sparsity projection guided by detection rates in the held-out tumor-derived reference population. Let \(\tilde{x}_{i}^{(H)}\) denote the decoded originating seed state after separate sparsity projection guided by the detection-rate structure of the corresponding held-out healthy-like seed subset. The pipeline-defined projected expression-space seed distance was quantified as
\[
d_{\mathrm{seed}}^{X}(\hat{x}_{i,b},x_i)
=
\left\|
\tilde{x}_{i,b}^{(T)}
-
\tilde{x}_{i}^{(H)}
\right\|_2.
\]
This quantity is interpreted as a reference-conditioned post hoc distance measure and not as distance in an unmodified common decoded expression space.

Because tumor-associated candidate acquisition is expected to move generated profiles along the observed healthy-to-tumor reference direction, projected seed distance alone cannot distinguish movement along this direction from remaining seed-conditioned structure. We therefore additionally evaluated distances after removing a held-out-reference-derived healthy-to-tumor axis.

Let \(\mu_H\) and \(\mu_T\) denote the mean decoded expression vectors of healthy-like and tumor-derived epithelial reference cells, respectively, from the held-out evaluation cohort. The unit healthy-to-tumor reference axis and its midpoint were defined as
\[
a_X =
\frac{\mu_T-\mu_H}
{\left\|\mu_T-\mu_H\right\|_2},
\qquad
c_X =
\frac{\mu_H+\mu_T}{2}.
\]
For a projected expression-space profile \(\tilde{x}\), its residualized representation was defined as
\[
R_X(\tilde{x})
=
(\tilde{x}-c_X)
-
\left[
(\tilde{x}-c_X)^\top a_X
\right] a_X.
\]
Residual expression-space seed distance was then defined as
\[
d_{\mathrm{resid}}^{X}(\hat{x}_{i,b},x_i)
=
\left\|
R_X(\tilde{x}_{i,b}^{(T)})
-
R_X(\tilde{x}_{i}^{(H)})
\right\|_2.
\]

For representative-candidate analyses, each retained representative candidate was compared with its own originating seed and with randomly reassigned seeds. For \(P=1000\) random seed permutations, the residual seed-dependence ratio was defined as
\[
\rho_{\mathrm{perm}}
=
\frac{
\overline{d}_{\mathrm{own}}
}{
\frac{1}{P}
\sum_{p=1}^{P}
\overline{d}_{\pi_p}
},
\]
where \(\overline{d}_{\mathrm{own}}\) denotes the mean residual distance between representative candidates and their originating seeds, and \(\overline{d}_{\pi_p}\) denotes the mean residual distance after permutation \(\pi_p\) of the seed assignments. Values below one indicate closer residual correspondence to the originating seed than expected under random seed reassignment.

The empirical one-sided permutation probability was computed as
\[
p_{\mathrm{perm}}
=
\frac{1}{P}
\sum_{p=1}^{P}
\mathbb{I}
\left(
\overline{d}_{\pi_p}
\leq
\overline{d}_{\mathrm{own}}
\right).
\]
These residual seed-distance summaries are interpreted as reference-conditioned descriptive measures and not as independently estimated predictive-performance measures.

\subsection{Residual candidate-cloud seed structure}

Because CancerZigZag generates stochastic candidate clouds rather than a single deterministic output per seed, we evaluated whether candidates generated from the same originating seed remained more similar to one another than candidates generated from different seeds after removal of the held-out-reference-derived healthy-to-tumor axis.

Let \(R_X(\tilde{x}_{i,b}^{(T)})\) denote the residualized projected expression-space representation of candidate realization \(b\) generated from seed \(i\). We estimated the mean within-seed residual distance as
\[
D_{\mathrm{within}}^{X,\mathrm{resid}}
=
\frac{1}{|\mathcal{P}_{\mathrm{within}}|}
\sum_{(i,b,b') \in \mathcal{P}_{\mathrm{within}}}
\left\|
R_X(\tilde{x}_{i,b}^{(T)})
-
R_X(\tilde{x}_{i,b'}^{(T)})
\right\|_2,
\qquad b \neq b',
\]
where \(\mathcal{P}_{\mathrm{within}}\) is a set of randomly sampled candidate pairs originating from the same seed. Analogously, the mean between-seed residual distance was estimated as
\[
D_{\mathrm{between}}^{X,\mathrm{resid}}
=
\frac{1}{|\mathcal{P}_{\mathrm{between}}|}
\sum_{(i,j,b,b') \in \mathcal{P}_{\mathrm{between}}}
\left\|
R_X(\tilde{x}_{i,b}^{(T)})
-
R_X(\tilde{x}_{j,b'}^{(T)})
\right\|_2,
\qquad i \neq j.
\]

Residual candidate-cloud seed structure was summarized by
\[
\rho_{\mathrm{cloud}}
=
\frac{
D_{\mathrm{within}}^{X,\mathrm{resid}}
}{
D_{\mathrm{between}}^{X,\mathrm{resid}}
}.
\]
Values of \(\rho_{\mathrm{cloud}}<1\) indicate that candidates generated from the same seed are, on average, more compact than candidates generated from different seeds after removal of the held-out-reference-derived healthy-to-tumor axis. Values close to \(1\) indicate weak or absent residual seed-conditioned structure at the level of the complete candidate cloud.

For each reported CancerZigZag configuration, within-seed and between-seed residual expression-space distances were estimated from 20,000 randomly sampled candidate pairs. These summaries are interpreted as reference-conditioned post hoc descriptions of candidate-cloud organization.

\subsection{Representative-candidate transcriptional-shift concordance}

To assess whether representative tumor-associated candidates exhibited transcriptional shifts in directions similar to those observed between real held-out tumor-derived and healthy-like epithelial cells, we performed a complementary gene-level concordance analysis.
Expression-space shifts were evaluated in the pipeline-defined reference-conditioned representation after sparsity projection of decoded generated candidates and corresponding seed representations as described above.

Let \(H_{\mathrm{test}}\) and \(T_{\mathrm{test}}\) denote the held-out healthy-like and tumor-derived epithelial reference populations, respectively. For each gene \(g\), the real held-out tumor-associated expression shift was defined as
\[
\Delta_g^{\mathrm{real}}
=
\overline{x}_{g,T_{\mathrm{test}}}
-
\overline{x}_{g,H_{\mathrm{test}}}.
\]

For each evaluated seed \(x_i\), let \(\hat{x}_{i,*}\) denote the representative candidate retained under the predefined tumor-probability and seed-distance rule. The representative-candidate expression shift was defined as
\[
\Delta_g^{\mathrm{CZ}}
=
\frac{1}{N}
\sum_{i=1}^{N}
\left(
\hat{x}_{i,*,g}
-
x_{i,g}
\right),
\]
where \(N\) denotes the number of evaluated seeds.

Concordance between generated and real tumor-associated shifts was quantified using Pearson and Spearman correlations between \(\Delta^{\mathrm{CZ}}\) and \(\Delta^{\mathrm{real}}\) across aligned genes. In addition, overlap was computed between the 200 genes showing the largest absolute expression shifts in the generated and real contrasts.

Pathway-level directional agreement was additionally evaluated using the 50 gene sets of the MSigDB Hallmark collection. A Hallmark gene set was counted as directionally agreeing when the representative-candidate shift and the held-out tumor-derived versus healthy-like reference shift had the same sign.

This analysis concerns one representative candidate per seed and is complementary to the primary candidate-cloud evaluation. It evaluates whether representative CancerZigZag outputs show transcriptional displacement in directions consistent with observed tumor-associated differences in held-out reference populations. It is not a differential-expression significance test, does not imply exact reconstruction of real tumor cells, and does not imply temporal or causal progression.

\subsection{Targeted falsification controls and multi-round ablation}

Because the present study focuses on the specific combination of unpaired seed initialization, stochastic candidate-cloud generation, and residual seed-dependence evaluation, we used targeted falsification controls and a multi-round ablation rather than treating existing methods as directly equivalent competitors. These analyses were designed to test whether the observed CancerZigZag behaviour could be explained by simpler alternative mechanisms.
These analyses were performed for the reported main configuration of each cancer context.

The setting $r=1$ served as a direct ablation of repeated ZigZag cycling. This
condition applies a single partial-noising/reverse-diffusion cycle using the
same perturbation depth $t$ as the corresponding multi-round CancerZigZag
configuration. It therefore tests whether the properties of multi-round
CancerZigZag candidate clouds can be explained by a single diffusion-editing
cycle.

A Gaussian matched-perturbation control was used to test whether local random
variation around each seed was sufficient to produce comparable
tumor-associated candidate states. In this control, candidates were generated
by adding Gaussian perturbations to the seed latent representations using a
perturbation scale matched to the displacement observed in the corresponding
CancerZigZag analysis.

Linear tumor-mode controls were used to test whether CancerZigZag candidates
were reducible to direct linear displacement toward tumor-associated reference
states.
Tumor-derived latent representations from the corresponding training
partition were summarized by eight tumor-cluster centroids obtained by
$k$-means clustering.
For each healthy-like
seed latent state $z_i$, linear-control candidates were generated by
interpolation toward the respective centroid:
\[
z_{i,c}^{(\alpha)}
=
z_i
+
\alpha
\left(
\mu_{T,c}^{Z}-z_i
\right),
\]
where $\mu_{T,c}^{Z}$ denotes one of the tumor-cluster centroids and
\[
\alpha \in
\{0.0,0.1,0.2,0.35,0.5,0.75,1.0,1.25,1.5\}.
\]
These linear controls were not interpreted as equivalent generative
competitors; they were designed specifically to test whether CancerZigZag
outputs could be explained by direct movement toward predefined tumor modes.

For each CancerZigZag candidate $\hat{z}_{i,b}$, linear explainability was
quantified by its shortest distance to any line segment connecting its
originating seed $z_i$ to a tumor-cluster centroid:
\[
d_{\mathrm{line}}(\hat{z}_{i,b})
=
\min_{c}
\frac{
\operatorname{dist}
\left(
\hat{z}_{i,b},
\operatorname{segment}
\left[
z_i,\mu_{T,c}^{Z}
\right]
\right)
}{
\left\|
\hat{z}_{i,b}-z_i
\right\|_2 + 10^{-9}
}.
\]
Values close to zero indicate that a candidate is well explained by simple
linear movement toward a tumor-cluster centroid. Candidates satisfying
\[
d_{\mathrm{line}}(\hat{z}_{i,b}) \leq 0.10
\]
were designated as linearly explainable under this control. Linear
explainability was evaluated for the complete generated candidate cloud, with
representative-candidate analyses reported only as supplementary summaries
where applicable.

For CancerZigZag and the $r=1$ ablation, stored intermediate latent states were
used to quantify generated-path geometry. Let
$z_{i,b}^{(0)},\ldots,z_{i,b}^{(L)}$ denote the stored latent states of one
generated path. The path-over-chord ratio was defined as
\[
\rho_{\mathrm{path}}
=
\frac{
\sum_{\ell=1}^{L}
\left\|
z_{i,b}^{(\ell)}
-
z_{i,b}^{(\ell-1)}
\right\|_2
}{
\left\|
z_{i,b}^{(L)}
-
z_{i,b}^{(0)}
\right\|_2
+
10^{-9}
}.
\]
A value close to one indicates an approximately straight generated path,
whereas larger values indicate non-straight movement through latent space. 

Together, these controls test whether CancerZigZag candidate clouds are
reducible to local random perturbation, a single diffusion-editing cycle, or
direct linear movement toward predefined tumor-associated reference modes.
They are interpreted as targeted falsification analyses rather than as a
performance ranking of equivalent generative methods.

\section{Results}\label{sec2}

\subsection{CancerZigZag generates seed-initialized candidate clouds from unpaired data}

We evaluated CancerZigZag across four epithelial cancer contexts: colorectal cancer (CRC), breast cancer (BRCA), lung cancer (LC), and renal cell carcinoma (RCC). In each context, healthy-like and tumor-derived epithelial cells were treated as unpaired source and reference populations. The diffusion model was trained exclusively on tumor-derived cells from the patient-wise training cohort, whereas healthy-like cells from held-out patients were used only as seed inputs during generation. For each held-out healthy-like seed cell, CancerZigZag generated a cloud of stochastic candidate states through repeated cycles of partial latent-space perturbation and reverse diffusion using the corresponding tumor-trained diffusion model (Fig.~\ref{fig:main_figure}).

CancerZigZag does not assign a unique tumor counterpart to an individual healthy-like seed cell. Its primary output is the complete candidate cloud generated from each seed. We therefore evaluated the generated outputs at complementary levels. Candidate-cloud analyses were used to assess tumor-associated candidate acquisition and residual seed-conditioned structure at the level of the stochastic output distribution. Representative-candidate summaries were used only as secondary, explicitly labelled summaries under the predefined tumor-probability and seed-distance rule. Stored generated paths were analysed descriptively to characterize properties of generated exploration paths, without interpreting these paths as temporal progression, lineage reconstruction, or deterministic cell fates.

Across the four cancer contexts, the reported CancerZigZag configurations generated candidate clouds containing candidates that reached the predefined post hoc reference-classifier threshold for every held-out seed included in the detailed downstream analyses. These candidates were obtained without paired healthy--tumor observations, conditional diffusion, classifier guidance, or supervised transition objectives during generation. At the same time, residual seed-conditioned structure varied across cancer contexts, indicating that tumor-associated candidate acquisition and retention of seed-specific structure were not uniformly coupled across datasets.

\subsection{Systematic parameter grids identify dataset-specific reported configurations}

We systematically evaluated CancerZigZag across a grid of the number of consecutive ZigZag cycles (r) and the perturbation depth (t), using
\[
r \in \{1,10,25,50,75,100\},
\qquad
t \in \{1,10,25,50,75,100\}.
\]
This resulted in 36 evaluated configurations per cancer context and 144 cancer-context/configuration combinations in total. The grid was used to assess how repeated tumor-trained diffusion cycles and perturbation depth affected tumor-associated candidate acquisition and residual seed-conditioned structure under a fixed candidate sampling budget.

The single-cycle configuration (r=1) provided a direct ablation of repeated ZigZag cycling. In contrast, configurations with (r>1) evaluated whether repeated partial perturbation and reverse diffusion enabled candidate states that were not obtained as robustly through a single diffusion-editing cycle. Across the evaluated grids, the relationship between post hoc tumor-association scoring and residual candidate-cloud structure differed by cancer context, motivating the selection of one reported reference-informed configuration per dataset for subsequent detailed analyses.

\begin{figure}[t]
\centering
\includegraphics[width=\linewidth]{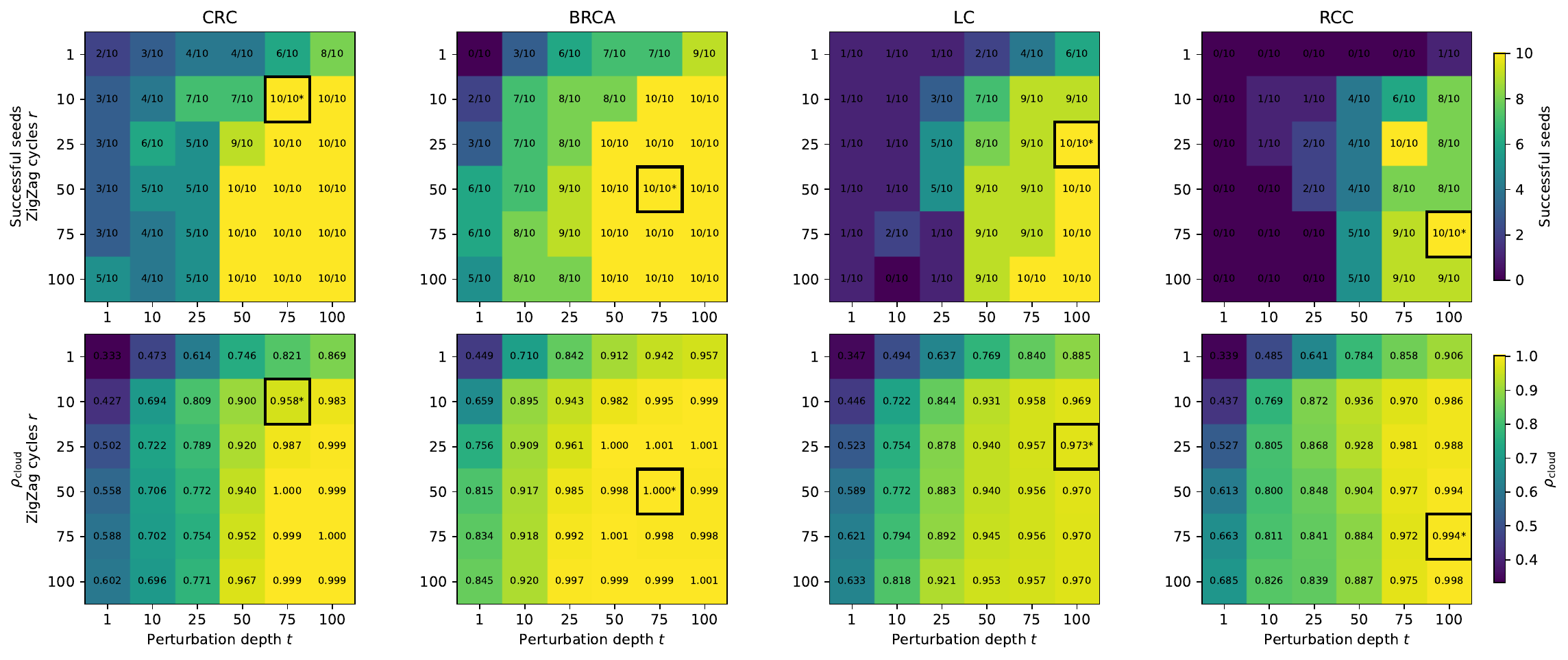}
\caption{\textbf{Reference-informed parameter landscapes for tumor-associated candidate acquisition and residual candidate-cloud structure.}
CancerZigZag was evaluated over the systematic parameter grid
\(r \in {1,10,25,50,75,100}\) and
\(t \in {1,10,25,50,75,100}\) separately for CRC, BRCA, LC, and RCC.
The upper row reports the number of held-out healthy-like seeds, out of ten evaluated seeds per configuration, for which the retained representative candidate reached the predefined classifier-estimated tumor-probability threshold of \(p_{\mathrm{tum}} \geq 0.7\).
The lower row reports residual candidate-cloud seed structure, quantified as the within-seed versus between-seed residual expression-space distance ratio \(\rho_{\mathrm{cloud}}\); values below (1) indicate residual seed-conditioned organization after removal of the held-out-reference-derived healthy-to-tumor axis.
Outlined cells marked with an asterisk indicate the reference-informed configurations selected for detailed downstream analyses.
The parameter landscapes were used for exploratory operating-regime identification rather than for independent predictive-performance estimation.}
\label{fig:parameter_grid_primary}
\end{figure}

The parameter landscapes revealed context-dependent relationships between tumor-associated candidate acquisition and residual candidate-cloud organization (Fig.~\ref{fig:parameter_grid_primary}). 
In CRC and BRCA, multiple configurations with moderate to high perturbation depths yielded threshold-reaching representative candidates for all ten evaluated seeds, whereas LC required higher perturbation depths and RCC showed the most restricted success region. The reference-informed configurations selected for detailed downstream analyses were (r=10, t=75) for CRC, (r=50, t=75) for BRCA, (r=25, t=100) for LC, and (r=75, t=100) for RCC. Each selected configuration yielded threshold-reaching representative candidates for all ten evaluated seeds. At the complete candidate-cloud level, residual within-seed organization was modest in CRC and LC and close to absent in the selected BRCA and RCC configurations, indicating that successful tumor-associated candidate acquisition did not necessarily imply persistent seed-conditioned organization of the full generated cloud.

\subsection{Representative candidates provide complementary tumor-associated summaries across cancer contexts}

Because the complete candidate cloud constitutes the primary output of CancerZigZag, we used representative candidates only as complementary summaries of each selected configuration. For each held-out healthy-like seed, candidates with a classifier-estimated tumor probability of at least (0.7) were designated as successful candidates. Among these successful candidates, the candidate with the smallest pipeline-defined projected expression-space distance to the originating seed was retained as the representative candidate. If no candidate reached the threshold, the candidate with the maximal tumor logit was retained as a fallback.

For all four selected configurations, each of the ten held-out seeds included in the detailed downstream analyses yielded at least one candidate reaching the predefined classifier-estimated tumor-probability threshold (Table~\ref{tab:zigzag_performance}). Mean classifier-estimated tumor probabilities of the retained representative candidates ranged from (0.779) in LC to (0.813) in RCC, with corresponding values of (0.804) in CRC and (0.803) in BRCA. These values quantify post hoc classification toward the observed tumor-derived evaluation reference population after the pipeline-defined expression-space sparsity projection. Importantly, this classifier was applied only during evaluation and post hoc summarization and did not guide CancerZigZag generation.

Representative candidates differed in their pipeline-defined projected expression-space distance from their originating seeds. Mean projected seed distances were (39.62) in LC, (40.52) in CRC, (43.26) in RCC, and (48.06) in BRCA. Because these expression-space distances are reference-conditioned post hoc summaries, they are not interpreted as direct measures of preserved cell identity. We therefore evaluated residual seed dependence and complete candidate-cloud organization after removal of the held-out-reference-derived healthy-to-tumor axis component.

\begin{table}[t]
\centering
\small
\caption{\textbf{Complementary representative-candidate and candidate-cloud summaries for the selected CancerZigZag configurations.}
For each seed, a representative candidate was selected by retaining the candidate with minimal projected expression-space distance among candidates reaching the predefined classifier-estimated tumor-probability threshold of \(p_{\mathrm{tum}} \geq 0.7\).
Success indicates the number of evaluated held-out seeds whose retained representative candidate reached this threshold.
Residual representative anchoring is summarized by \(\rho_{\mathrm{perm}}\), the ratio between own-seed and permuted-seed residual distances.
Residual cloud structure is summarized by \(\rho_{\mathrm{cloud}}\), the ratio between within-seed and between-seed residual candidate-cloud distances.
For both ratios, values below \(1\) indicate residual seed-conditioned structure in the corresponding reference-conditioned analysis.}
\label{tab:zigzag_performance}
\begin{tabular}{@{}lccccccc@{}}
\toprule
Dataset & \((r,t)\) & Success & \(\bar{p}_{\mathrm{tum}}\) & \(\bar{d}_{\mathrm{seed}}^{X}\) & \(\rho_{\mathrm{perm}}\) & \(p_{\mathrm{perm}}\) & \(\rho_{\mathrm{cloud}}\) \\
\midrule
CRC  & \((10,75)\)  & \(10/10\) & \(0.804\) & \(40.52\) & \(0.859\) & \(<0.001\) & \(0.958\) \\
BRCA & \((50,75)\)  & \(10/10\) & \(0.803\) & \(48.06\) & \(0.976\) & \(0.028\)  & \(1.000\) \\
LC   & \((25,100)\) & \(10/10\) & \(0.779\) & \(39.62\) & \(0.942\) & \(<0.001\) & \(0.973\) \\
RCC  & \((75,100)\) & \(10/10\) & \(0.813\) & \(43.26\) & \(0.965\) & \(0.002\)  & \(0.994\) \\
\bottomrule
\end{tabular}
\end{table}

\subsection{Residual representative-candidate anchoring is strongest in CRC and LC, whereas cloud-level structure is weak}

We next assessed whether CancerZigZag outputs retained measurable seed-conditioned structure beyond their position along the held-out-reference-derived healthy-to-tumor axis. Residual representative-candidate anchoring was summarized as the ratio between residual distance to the originating seed and the corresponding mean residual distance after random seed reassignment. Values below \(1\) indicate closer residual correspondence to the originating seed.

Residual representative-candidate anchoring was most evident in CRC, with an own-seed versus permuted-seed ratio of \(0.859\), followed by LC with a ratio of \(0.942\). RCC and BRCA yielded ratios closer to one (\(0.965\) and \(0.976\), respectively), indicating more limited effect sizes in the reported configurations. In the corresponding permutation analyses, none of 1,000 random seed reassignments yielded a residual mean distance as small as that observed for the true candidate--seed pairing in CRC or LC. For BRCA, 28 of 1,000 permutations met this criterion (\(p_{\mathrm{perm}}=0.028\)), whereas 2 of 1,000 permutations did so for RCC (\(p_{\mathrm{perm}}=0.002\)).

Because representative-candidate summaries capture only one retained output per seed, we additionally evaluated organization of the complete candidate clouds. Residual within-seed versus between-seed distance ratios were close to one in every cancer context: 0.958 in CRC, 0.973 in LC, 0.994 in RCC, and 1.000 in BRCA. These values indicate, at most, weak residual cloud-level organization under the reported configurations. Accordingly, the supported finding is that representative candidates retain residual seed dependence most clearly in CRC and LC, rather than that complete CancerZigZag candidate clouds remain strongly seed-conditioned after tumor-associated candidate acquisition.

\subsection{Targeted controls support multi-round candidate discovery and nonlinear generated paths}

We next used targeted controls to test specific simpler explanations for CancerZigZag outputs. These controls were not treated as equivalent generative competitors or as a conventional benchmark leaderboard. Instead, they tested whether threshold-reaching candidate discovery could be achieved by a single partial-noising/reverse-diffusion cycle or by matched local Gaussian perturbation, and whether CancerZigZag paths were reducible to direct linear movement toward explicit tumor-associated latent centroids.

Across all four cancer contexts, the reported multi-round CancerZigZag configurations yielded threshold-reaching candidates for all ten downstream-evaluated seeds. The corresponding single-cycle ablation yielded successful candidates for \(6/10\) CRC seeds, \(7/10\) BRCA seeds, \(6/10\) LC seeds, and only \(1/10\) RCC seeds. Matched Gaussian perturbation yielded successful candidates for \(7/10\), \(8/10\), \(8/10\), and \(5/10\) seeds in CRC, BRCA, LC, and RCC, respectively. Thus, in the evaluated downstream seed subsets, multi-round CancerZigZag produced threshold-reaching candidates more consistently than either the single-cycle or matched Gaussian control (Table~\ref{tab:targeted_controls}).

Direct interpolation toward tumor-cluster centroids also yielded threshold-reaching candidates for all evaluated seeds. This control explicitly uses tumor-derived reference centroids and was therefore interpreted as a targeted linear explainability test rather than as an equivalent unsupervised generative baseline. Under the predefined line-distance threshold \(d_{\mathrm{line}}\leq 0.10\), none of the CancerZigZag candidates were linearly explainable by direct seed-to-tumor-cluster-centroid movement in any of the four cancer contexts. Stored CancerZigZag paths were also strongly non-straight in latent space, with mean path-over-chord ratios of \(7.202\) in CRC, \(34.505\) in BRCA, \(17.050\) in LC, and \(44.889\) in RCC, whereas the corresponding single-cycle paths had path-over-chord ratios of \(1.000\).

\begin{table}[t]
\centering
\small
\caption{\textbf{Targeted controls for candidate discovery and generated-path geometry.}
Success denotes the number of ten evaluated seeds with at least one candidate reaching the predefined classifier-estimated tumor-probability threshold. The cluster-linear control is an explicit reference-mode interpolation control rather than an equivalent generative baseline. The column ``Linear'' reports the fraction of CancerZigZag candidates classified as linearly explainable under $d_{\mathrm{line}} \leq 0.10$.}
\label{tab:targeted_controls}
\begin{tabular}{@{}lcccccc@{}}
\toprule
Dataset & CZ & $r=1$ & Gaussian & Cluster-linear & Linear & $\rho_{\mathrm{path}}$ \\
\midrule
CRC  & $10/10$ & $6/10$ & $7/10$ & $10/10$ & $0.000$ & $7.202$ \\
BRCA & $10/10$ & $7/10$ & $8/10$ & $10/10$ & $0.000$ & $34.505$ \\
LC   & $10/10$ & $6/10$ & $8/10$ & $10/10$ & $0.000$ & $17.050$ \\
RCC  & $10/10$ & $1/10$ & $5/10$ & $10/10$ & $0.000$ & $44.889$ \\
\bottomrule
\end{tabular}
\end{table}

Together, these targeted controls indicate that multi-round CancerZigZag improves threshold-reaching candidate discovery over single-cycle and matched Gaussian perturbation controls in the analysed seed subsets, while generating nonlinear latent-space paths that are not reducible to direct interpolation toward the evaluated tumor-cluster reference modes.

\subsection{Representative candidates show context-dependent transcriptional-shift concordance with held-out references}

We next evaluated whether representative CancerZigZag candidates exhibited expression changes consistent with tumor-associated differences in the held-out evaluation references. These analyses used the pipeline-defined, reference-conditioned expression representation after sparsity projection of generated candidate profiles. For each cancer context, representative-candidate expression shifts relative to the corresponding originating seeds were compared with tumor-derived versus healthy-like reference shifts across aligned genes. This analysis quantifies gene-level shift concordance and is not a differential-expression significance test.

Gene-level transcriptional-shift concordance was positive in all four cancer contexts (Table~\ref{tab:transcriptional_concordance}). Pearson correlations were \(r=0.571\) in CRC, \(r=0.894\) in BRCA, \(r=0.588\) in LC, and \(r=0.747\) in RCC; corresponding Spearman correlations were \(\rho=0.487\), \(\rho=0.564\), \(\rho=0.454\), and \(\rho=0.645\), respectively. Overlap among the 200 genes with the largest absolute shifts was partial, ranging from \(25/200\) genes in LC to \(90/200\) genes in RCC. These results support partial, context-dependent agreement in gene-level expression-shift patterns rather than exact reconstruction of observed tumor-derived profiles.

We additionally summarized pathway-level direction agreement using the 50 gene sets of the MSigDB Hallmark collection. A Hallmark gene set was counted as directionally agreeing when its representative-candidate shift and its held-out tumor-derived versus healthy-like reference shift had the same sign. Direction agreement was observed for \(35/50\) Hallmark gene sets in CRC, \(35/50\) in BRCA, \(47/50\) in LC, and \(50/50\) in RCC.

\begin{table}[t]
\centering
\small
\caption{\textbf{Reference-conditioned transcriptional-shift concordance of representative CancerZigZag candidates.}
Gene-level correlations compare representative-candidate expression shifts relative to originating seeds with held-out tumor-derived versus healthy-like reference shifts. Top-200 overlap denotes shared genes among the 200 largest absolute shifts. Hallmark direction agreement reports the number of the 50 MSigDB Hallmark gene sets with the same shift direction in generated and held-out reference contrasts.}
\label{tab:transcriptional_concordance}
\begin{tabular}{@{}lcccc@{}}
\toprule
Dataset & Pearson $r$ & Spearman $\rho$ & Top-200 overlap & Hallmark agreement \\
\midrule
CRC  & $0.571$ & $0.487$ & $44/200$ & $35/50$ \\
BRCA & $0.894$ & $0.564$ & $47/200$ & $35/50$ \\
LC   & $0.588$ & $0.454$ & $25/200$ & $47/50$ \\
RCC  & $0.747$ & $0.645$ & $90/200$ & $50/50$ \\
\bottomrule
\end{tabular}
\end{table}

These results provide reference-conditioned biological characterization of representative CancerZigZag outputs beyond the scalar classifier score. Because candidate evaluation and expression-space representation depend on the held-out reference population and the pipeline-defined sparsity projection, this concordance is not interpreted as independent validation, exact reconstruction of tumor states, temporal progression, or causal tumor-development modelling.

\section{Discussion}\label{sec12}

In this study, we introduce CancerZigZag, a diffusion-based framework for seed-initialized exploration of tumor-associated single-cell candidate states from unpaired epithelial cell populations. The central methodological idea is to separate the learning of a tumor-associated generative landscape from the initialization of candidate generation: a diffusion model is trained exclusively on tumor-derived epithelial cells, whereas held-out healthy-like cells are introduced only as starting states for iterative stochastic exploration. In contrast to deterministic state-translation approaches, CancerZigZag generates a candidate cloud for each seed, allowing tumor-associated scoring and residual seed dependence to be examined as distinct properties of the generated outputs.

Across the four analysed cancer contexts, the reported operating configurations generated candidate clouds containing candidates that reached the predefined reference-classifier threshold for each evaluated held-out seed. These results show that repeated application of a tumor-trained diffusion operator can generate candidate states classified toward held-out tumor-derived reference populations without using paired healthy--tumor observations or classifier guidance during generation. However, because the reported configurations were identified from reference-informed parameter landscapes, these findings should be interpreted as exploratory characterization of candidate-generation regimes rather than as estimates of independently preselected predictive performance.

A central observation of the study is that tumor-associated candidate acquisition and residual seed dependence do not coincide uniformly across cancer contexts. CRC showed the clearest residual representative-candidate anchoring and the lowest residual candidate-cloud ratio, although cloud-level organization remained weak overall. LC retained more modest residual representative-candidate anchoring, whereas the reported BRCA and RCC settings showed only limited residual structure. By contrast, the reported RCC and BRCA configurations generated threshold-reaching candidates for all evaluated seeds but showed limited residual organization at the candidate-cloud level. These results indicate that seed initialization does not guarantee persistent seed-dependent structure after tumor-associated displacement. Instead, CancerZigZag exposes a context-dependent relationship between movement toward a tumor-associated reference distribution and retention of information associated with the originating seed.

This interpretation is important for defining the scope of the method. CancerZigZag is not intended to reconstruct temporal disease progression, infer lineage relationships, or predict the future malignant state of an individual healthy-like cell. The generated candidates are model-produced states explored from a seed within a tumor-trained latent landscape. Accordingly, residual seed dependence is interpreted as a measurable property of the generated candidate distribution, not as evidence that a generated tumor-associated state is the biological descendant or unique counterfactual outcome of its originating seed. The framework is therefore best understood as a hypothesis-generating approach for exploring tumor-associated candidate states under explicit evaluation criteria.

The transcriptional-shift analyses provide an additional descriptive layer beyond reference-classifier scoring. Representative candidates exhibited directional agreement with expression shifts observed between held-out healthy-like and tumor-derived reference populations, and pathway-level analyses were used to assess whether selected outputs reflected coordinated tumor-associated programmes. These analyses are relevant because they examine whether generated candidates display structured expression changes rather than merely obtaining elevated classifier scores. Nevertheless, they do not establish exact reconstruction of observed tumor cells, recovery of causal regulatory mechanisms, or correspondence to clinically defined tumor states. Their interpretation should remain limited to directional concordance with held-out tumor-associated reference patterns.

Several limitations define the appropriate use of CancerZigZag. First, the candidate space generated by the model is constrained by the representation learned by the VAE and by the diversity, coverage, and potential biases of the tumor-derived training population. Consequently, the framework cannot be assumed to recover intermediate, premalignant, rare, or clinically relevant tumor-associated states that are insufficiently represented in the training data. Second, the present evaluation is reference-conditioned: held-out reference populations are used for post hoc tumor-associated scoring, sparsity adjustment, residualization, and transcriptional comparison. The reported analyses therefore characterize how generated candidates relate to observed reference populations, rather than providing an independent prediction of biological malignancy or disease progression. Third, CancerZigZag does not incorporate regulatory priors, mechanistic constraints, treatment-response information, or clinical annotations during generation. Extensions incorporating external biological references, perturbation data, or validated regulatory constraints will be required before generated candidates can be interpreted in mechanistic or translational terms.

Taken together, CancerZigZag provides a framework for exploratory generation and evaluation of tumor-associated candidate clouds in settings where paired or longitudinal single-cell observations are unavailable. Its contribution is not an identity-preserving conversion from healthy-like to tumor cells, but the ability to initialize stochastic exploration from individual healthy-like cells and to quantify how tumor-associated scoring and residual seed dependence vary across generated candidate distributions and biological contexts. This framework may provide a useful basis for future studies of unpaired cell-state exploration, provided that generated candidates are treated as hypotheses for downstream validation rather than as predictions of true cellular trajectories.

\section{Conclusion}\label{sec13}

We present CancerZigZag, a diffusion-based framework for seed-initialized generation of tumor-associated single-cell candidate clouds from unpaired epithelial cell populations. By applying repeated partial perturbation and reverse diffusion using models trained exclusively on tumor-derived cells, CancerZigZag enables exploratory generation from healthy-like seed cells without paired healthy--tumor observations or guidance during generation.

Across the reported operating configurations, generated candidate clouds contained candidates classified toward held-out tumor-derived reference populations for each evaluated seed, while residual seed-dependent organization varied substantially across cancer contexts. These findings position CancerZigZag not as a model of deterministic cellular transformation or disease progression, but as a framework for reference-informed exploration of tumor-associated candidate states and the relationship between tumor-associated displacement and residual seed dependence in unpaired single-cell data.

\backmatter

\section*{Declarations}

\subsection*{Funding}
This work was funded by the European Union under the Horizon Europe grant 101186829 (CancerScan).

\subsection*{Competing interests}
The authors have no relevant financial or non-financial interests to disclose.

\subsection*{Ethics approval and consent to participate}
Not applicable. This study is a secondary computational analysis of previously collected, de-identified single-cell transcriptomic data. No new human participants were recruited, no new biological samples were collected, and no individual-level identifying information was accessed for this study.

\subsection*{Consent for publication}
Not applicable. This manuscript does not contain identifiable individual participant data, images, or other information requiring consent for publication.

\subsection*{Data availability}
The single-cell transcriptomic data analysed in this study are publicly available from the pan-cancer tumor--normal single-cell atlas by Kang et al.~\cite{Kang2024}. The normalized single-cell RNA-sequencing expression matrices used for the CRC, BRCA, LC, and RCC analyses were obtained from the associated Zenodo repository: \url{https://doi.org/10.5281/zenodo.10651059}.

\subsection*{Materials availability}
Not applicable. No new biological materials were generated or used in this computational study beyond the previously collected single-cell transcriptomic datasets described above.

\subsection*{Code availability}
The code implementing CancerZigZag generation, candidate selection, targeted falsification analyses, trajectory analyses, and biological programme evaluation is available at \url{https://github.com/JHelge/CanzerZigZag}. The repository includes the scripts and configuration files required to reproduce the analyses reported in this manuscript.

\subsection*{Author contributions}
J.S. conceived the study, developed the methodology, implemented the computational framework, performed the analyses, interpreted the results, prepared the figures, and wrote the original manuscript draft. A.S. critically reviewed the methodology and interpretation of results and revised the manuscript for important intellectual content. All authors approved the final manuscript and agreed to be accountable for the work.

\bibliography{sn-bibliography}


\end{document}